\documentclass[aps,prl,twocolumn,superscriptaddress]{revtex4-1}
\usepackage{graphicx}
\usepackage{dcolumn}
\usepackage{bm}

\def\vv{{\bf v}}
\def\hvp{\hat{\bf p}}
\def\vare{\varepsilon}
\newcommand{\Del}{{\mbox{\footnotesize $\Delta$}}}
\def\un{\textmd{\scriptsize un}}

\begin{document}

\title{London penetration depth in Ba(Fe$_{1-x}$T$_x$)$_2$As$_2$ (T=Co, Ni)\\
superconductors irradiated with heavy ions}

\author{H.~Kim}
\affiliation{Ames Laboratory and Department of Physics \& Astronomy, Iowa State University, Ames, IA 50011}

\author{R.~T.~Gordon}
\affiliation{Ames Laboratory and Department of Physics \& Astronomy, Iowa State University, Ames, IA 50011}

\author{M.~A.~Tanatar}
\affiliation{Ames Laboratory and Department of Physics \& Astronomy, Iowa State University, Ames, IA 50011}

\author{J.~Hua}
\affiliation{Materials Science Division, Argonne National Laboratory, Argonne, Illinois 60439}

\author{U.~Welp}
\affiliation{Materials Science Division, Argonne National Laboratory, Argonne, Illinois 60439}

\author{W.~K.~Kwok}
\affiliation{Materials Science Division, Argonne National Laboratory, Argonne, Illinois 60439}

\author{N.~Ni}
\affiliation{Ames Laboratory and Department of Physics \& Astronomy, Iowa State University, Ames, IA 50011}

\author{S.~L.~Bud'ko}
\affiliation{Ames Laboratory and Department of Physics \& Astronomy, Iowa State University, Ames, IA 50011}

\author{P.~C.~Canfield}
\affiliation{Ames Laboratory and Department of Physics \& Astronomy, Iowa State University, Ames, IA 50011}

\author{A.~B.~Vorontsov}
\affiliation{Department of Physics, Montana State University, Bozeman, Montana, 59717}

\author{R.~Prozorov}
\email[Corresponding author: ]{prozorov@ameslab.gov}
\affiliation{Ames Laboratory and Department of Physics \& Astronomy, Iowa State University, Ames, IA 50011}

\date{8 April 2010}

\begin{abstract}
Irradiation with Pb ions was used to study the effect of disorder on the in-plane London penetration depth, $\lambda(T)$, in single crystals of Ba(Fe$_{1-x}$T$_x$)$_2$As$_2$ (T=Co, Ni). An increase of the irradiation dose results in a monotonic decrease of the superconducting transition temperature, $T_c$, without affecting much the transition width. In both Co and Ni doped systems we find a power-law behavior, $\Delta\lambda(T) \propto T^n$, with the exponent $n$ systematically decreasing with the increase of disorder. This observation, supported by the theoretical analysis, conclusively points to a nodeless $s^\pm$ state with pairbreaking impurity scattering (interband) with strength being intermediate between Born and unitary limits.
\end{abstract}


\maketitle

The mechanism of superconductivity in Fe-based superconductors \cite{Kamihara2008} is a focus of extensive experimental \cite{Ishida2009} and theoretical \cite{Chubukov2009,Mazin2009} efforts. Understanding the
superconducting gap structure is crucial for identifying the pairing mechanism. While most of the
experiments suggest that the general structure of the pairing state belongs to the most symmetric
$A_{1g}$ class, the multi-band nature of these materials allows for a number of possibilities, including
the conventional $s$-wave state, extended $s^\pm$-state with different signs of the order parameter on
two bands \cite{Mazin2008,Kuroki08,Barzykin08,Mazin2009} and states with highly anisotropic gaps and even
nodes \cite{SeoBernevig08,Maier09gap,Graser09,Chubukov09nodes,Thomale09nodes}.

The London penetration depth can be measured with great precision and its variation with temperature depends
sensitively on the gap structure. For $T \leq T_c/3$, a conventional isotropic $s$-wave gap $\Delta_0$
results in an exponential behavior, $\Delta\lambda(T) \propto \exp({-\Delta_0/T})$, which is preserved
even with the addition of non-magnetic impurities \cite{Tinkham}. Unconventional pairing states, on
the other hand, are susceptible to the presence of non-magnetic impurities
\cite{Chubukov2009,Bang2009,Vorontsov2009,Gordon2010}. In nodal $d$-wave superconductors, $\lambda(T)$ exhibits a power-law behavior, $\Delta\lambda(T) \propto T^n$, with the exponent $n$ \emph{increasing} from $n$=1 in the clean case to $n$=2 in the dirty limit \cite{Hirschfeld1993}. For the extended $s^\pm$ state, the \emph{opposite} trend is expected: $\Delta\lambda(T)$ is exponential in the clean limit, changing with disorder to a power-law, with $n$ as low as 1.6 \cite{Bang2009,Vorontsov2009}.

Experimentally, a power-law behavior with the exponent $2 \leq n < 3$ has been observed in most of the iron-based superconductors \cite{Gordon2009,Gordon2009a,Hashimoto2009a,Martin2009a,Martin2009,Martin2010,Luan2010,Kim2010}. This characteristic
exponent $n \sim$2 can be explained in both dirty $d$-wave and dirty $s^{\pm}$ scenarios. The disorder
is always present in the iron-based superconductors where doping (e.g., Co or Ni in this work) is
needed to induce superconductivity. One way to resolve this complication is to deliberately introduce
defects that do not contribute extra charge but rather only increase the scattering rate. Various ways
of controlling the scattering rate have been suggested in earlier studies, especially for the high-$T_c$ cuprates, and the effects have been examined by using transport and magnetic
measurements \cite{Jin1993,Blatter1994}. In particular, irradiation with heavy ions, besides producing efficient pinning centers, also significantly enhances scattering, as evident from the significant increase of the normal state resistivity \cite{Zhu1993,Woods1998} as well as the suppression of $T_c$.

In this work, we study the in-plane London penetration depth in single crystals of optimally Co-and
Ni-doped BaFe$_2$As$_2$ superconductors irradiated with 1.4 GeV $^{208}$Pb$^{56+}$ ions. In both systems
we find monotonic suppression of $T_c$ with the increase of the irradiation dose without notable transition broadening. The London penetration depth exhibits a power-law behavior, $\Delta\lambda(T)\propto A\, T^n$ ($2.2 < n < 2.8$), with the exponent $n$  \emph{decreasing} with the irradiation dose. This observation, supported by the theoretical analysis, provides the most convincing
case for the nodeless $s^{\pm}$ state with pairbreaking scattering (interband) of intermediate strength, between Born and unitary scattering limits.

Single crystals of Ba(Fe$_{1-x}$T$_x$)$_2$As$_2$ (T=Co, Ni denoted FeCo122 and FeNi122, respectively)
were grown out of FeAs flux using a high temperature solution growth technique
\cite{Canfield2009,Ni2008}. X-ray diffraction, resistivity, magnetization and wavelength dispersive
spectroscopy (WDS) elemental analysis have all shown good quality single crystals at the optimal dopings
with a small variation of the dopant concentration over the sample and sharp superconducting
transitions, $T_c$= 22.5 K for FeCo122 and 18.9 K for FeNi122 \cite{Canfield2009,Ni2008}. The in-plane
London penetration depth was measured by using the tunnel diode resonator technique
\cite{Degrift1975,Prozorov2000,Prozorov2000a,Prozorov2006}. The sample was placed with its
crystallographic $c$-axis parallel to a small excitation field, $H_{ac} \approx 20$ mOe. The shift of
the resonant frequency, $\Delta f(T)$, is proportional to the sample magnetic susceptibility, $\chi (T)$
via $\Delta f(T) = -G4\pi\chi (T)$. Here $G$ is a geometric calibration factor, $G=f_0V_s/2V_c(1-N)$,
where $N$ is the demagnetization factor, $V_s$ is the sample volume, and $V_c$ is the coil volume. The
calibration factor was determined from the full frequency change by physically pulling the sample out of
the coil. The magnetic susceptibility can be written in terms of $\lambda$ and the characteristic length
$R$, $4\pi\chi=(\lambda /R) \tanh (R/\lambda) -1$, from which $\Delta\lambda$ can be acquired
\cite{Prozorov2000}.

To examine the effect of irradiation, $\sim 2\times 0.5\times 0.02-0.05$ mm$^3$ single crystals were
selected and then cut into several pieces preserving the width and the thickness. We compare sets of
samples, where the samples in each set are parts of the same original large crystal. Several such sets
were prepared and a reference piece was kept unirradiated from each set. Irradiation with 1.4 GeV
$^{208}$Pb$^{56+}$ ions was performed at the Argonne Tandem Linear Accelerator System (ATLAS) with an
ion flux of $\sim 5\times 10^{11}$ ions$\cdot$s$^{-1}$$\cdot$m$^{-2}$. The actual total dose was
recorded in each run. The density of defects ($d$) created by the irradiation is usually expressed in
terms of the matching field, $B_\phi=\Phi_0 d$, which is obtained assuming one flux quanta,
$\Phi_0\approx 2.07\times 10^{-7}$ G$\cdot$cm$^2$ per ion track. Here we studied samples with
$B_\phi=0.5$, 1.0 and 2.0 T corresponding to $d=2.4\times 10^{10}$ cm$^{-2}$, $4.8\times 10^{10}$
cm$^{-2}$ and $9.7\times 10^{10}$ cm$^{-2}$. The sample thickness was chosen in the range of $\sim 20 -
50 \mu$m to be smaller than the ion penetration depth, $60 - 70~\mu$m. The same samples were studied by
magneto-optical imaging.  The strong Meissner screening and large uniform enhancement of pinning have
shown that the irradiation has produced uniformly distributed defects \cite{Prozorov2010}.

\begin{figure}[tb]
\includegraphics[width=1.0\linewidth]{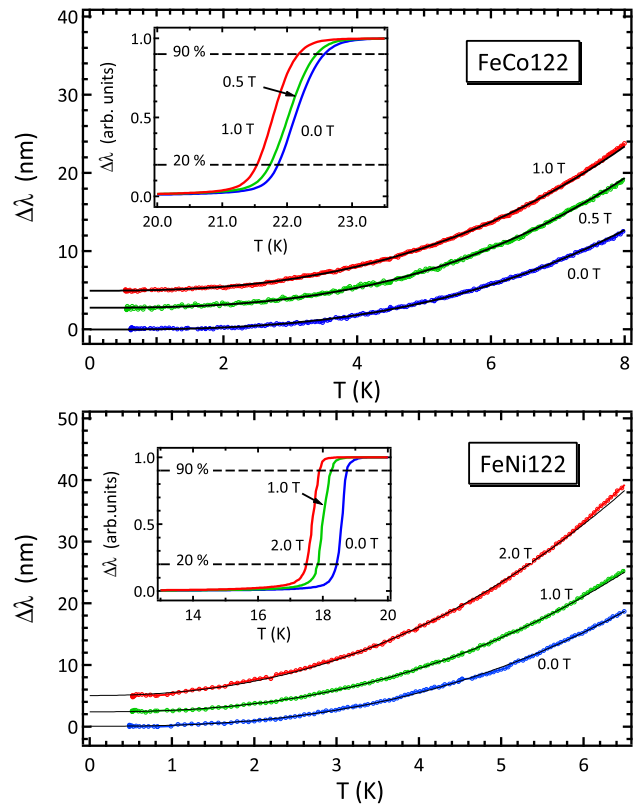}%
\caption{Variation of the in-plane London penetration depth, $\Delta\lambda(T)$, for
irradiated FeCo122 (top panel) and FeNi122 (bottom panel). The low-temperature variations are shown
in the main frame of each panel along with the best power-law fits. The curves are offset vertically
for clarity. The variations in the vicinity of $T_c$ are shown in the insets of each panel.}
\label{fig1}
\end{figure}

Figure \ref{fig1} shows $\Delta\lambda(T)$ for FeCo122 (top panel) and FeNi122 (bottom panel). The
low-temperature region up to $\approx T_c/3$ is shown in the main frame of each panel. Vertical offsets
were applied for clarity. The normalized penetration depths in the vicinity of $T_c$ are shown in the
inset of each panel to highlight the suppression of $T_c$ as the radiation dose increases. Whereas $T_c$
is clearly suppressed, the transition width remains nearly the same (see Fig.~\ref{fig3} below). All
samples exhibit a power-law variation of $\Delta\lambda(T)\propto T^n$ with $2.5< n <2.8$ up to $T_c/3$,
while the exponential fitting failed in all cases. The best fitting curves are shown by solid lines in
Fig.~\ref{fig1}. We note that the present set of FeCo122 samples exhibits higher exponents, $n$,
compared to previous works \cite{Gordon2009a,Luan2010}. This variation of $n$ is likely
due to disorder, as we clearly demonstrate in this work. Consequently, it is important to conduct a
comparison of radiation effects on the samples made from the \emph{same} large crystal. Magneto-optical
characterization has shown a homogeneous superconducting response \cite{Prozorov2010} and the widths of the superconducting transitions were much smaller than the absolute shift due to irradiation, see Fig.~\ref{fig3}. Therefore, it is very likely that the effects reported here are caused by the enhanced scattering induced by the heavy ion bombardment.

\begin{figure}
\includegraphics[width=1.0\linewidth]{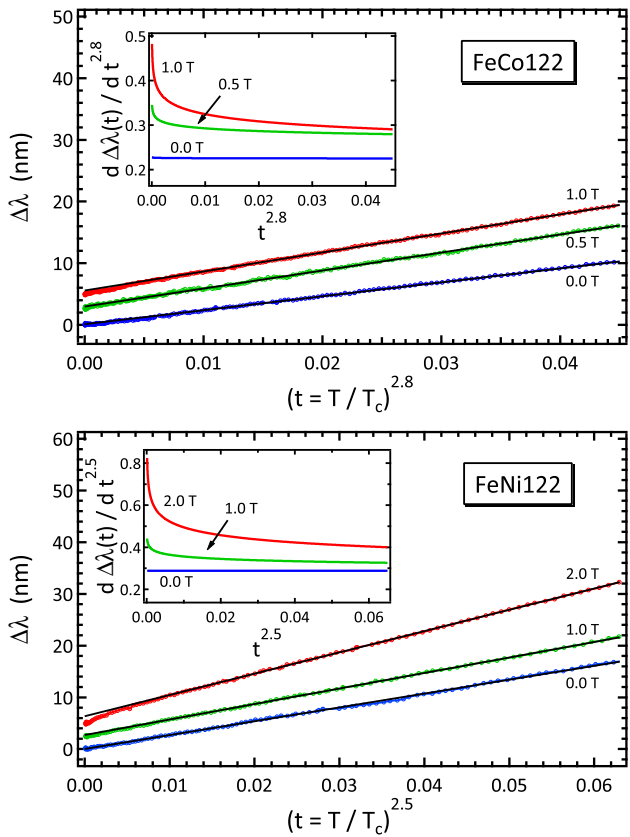}%
\caption{Detailed comparison of the functional form of $\Delta\lambda(T)$ for irradiated FeCo122
and FeNi122. In the main panels $\Delta\lambda(T)$ is plotted vs. $(t=T/T_c)^{n_0}$ with the
exponents $n$ taken from the best fits of \emph{unirradiated} samples: $n_0 =$ 2.8 and 2.5 for
FeCo122 and FeNi122, respectively (see Fig.\ref{fig1}). Apparently, irradiation causes
low-temperature deviations, which are better seen in the derivatives, $d \Delta\lambda(t)/dt^{n_0}$,
plotted in the insets.}
\label{fig2}
\end{figure}

To further analyze the power-law behavior and its variation with irradiation, we plot $\Delta\lambda$ as
a function of  $(t=T/T_c)^{n_0}$ in Fig.~\ref{fig2}, where the $n_0$ values for FeCo122 and FeNi122 were
chosen from the best power-law fits of the unirradiated samples (see Fig.~\ref{fig3}). While the data
for unirradiated samples appear as almost perfect straight lines showing robust power-law behavior, the
curves for irradiated samples show downturns at low temperatures indicating smaller exponents. This
observation, emphasized by the plots of the derivatives $d\Delta\lambda(t)/dt^{n_0}$ in the inset of
Fig.~\ref{fig2}, points to a significant change in the low-energy excitations with radiation.

\begin{figure}
\includegraphics[width=1.0\linewidth]{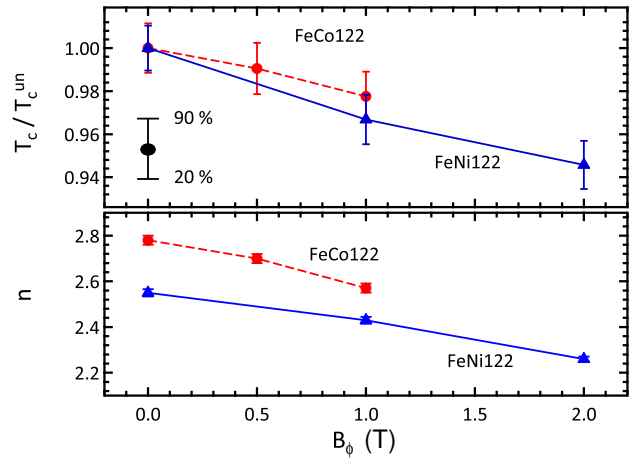}%
\caption{Top panel: The suppression of $T_c$ with disorder relative to
\emph{unirradiated} $T^{\un}_c$. The vertical bars denote the width of the transition and correspond to temperatures where the diamagnetic response changed from 90\% (onset) to 20\% (end of the transition),
see insets in Fig.~\ref{fig1}. Lower panel: exponent $n$ vs. $B_{\phi}$.}
\label{fig3}
\end{figure}

The variations of $T_c$ and $n$ upon irradiation are illustrated in
Fig.~\ref{fig3}. Dashed lines and circles show FeCo122, while solid lines and
triangles show FeNi122. The upper panel shows the variation of $T_c$ and the width of
the transition. Since $B_{\phi}$ is directly proportional to the area density of the ions, $d$, we can say that $T_c$ decreases roughly linearly with $d$. The same trend is evident for the exponent $n$ shown in the lower panel of Fig.~\ref{fig3}. The fitting pre-factor $A$ increases somewhat upon the increase of irradiation dose, but remains smaller than the value measured previously in unirradiated samples \cite{Gordon2009,Gordon2009a,Luan2010}.

The experimental results fit comfortably within the hypothesis of $s^\pm$ superconductivity with two isotropic gaps. The superfluid density in linear response is
\begin{equation}
\rho(T) =
\sum_{i=1,2} \pi T\sum_{\vare_m} N_{f,i} \int\limits_{FS_i} d\hvp
[\vv_{f,i}\otimes \vv_{f,i}]_{xx}
\frac{\tilde\Delta_i^2 }
{(\tilde\vare_m^2 + \tilde\Delta_i^2 )^{3/2} }
\end{equation}
\noindent where one sums over the contributions from the electron and hole bands;
$\vv_{f,i}$ and $N_{f,i}$ are the Fermi velocity and density of states in these
bands, taken to be equal for the calculations. Two order parameters $\Delta_{1,2}$
are computed self-consistently together with the $t$-matrix treatment of
impurity effects, which renormalize the Matsubara energies
$i\tilde\vare_m = i\vare_m - \Sigma_{imp,i}$
and the gaps $\tilde\Delta_i = \Delta_i + \Delta_{imp,i}$\cite{mishra09}.
Impurities are characterized by the strength of the potential for scattering
within each band, $v_{11}(=v_{22})$, given by the phase shift
$\delta = \tan^{-1}(\pi N_f v_{11})$, the ratio of potentials for inter-band and
intra-band hopping, $\delta v = v_{12}/v_{11}$, and the impurity
scattering rate $\Gamma = n_{imp}/\pi N_f$.

\begin{figure}[tb]
\includegraphics[width=1.0\linewidth]{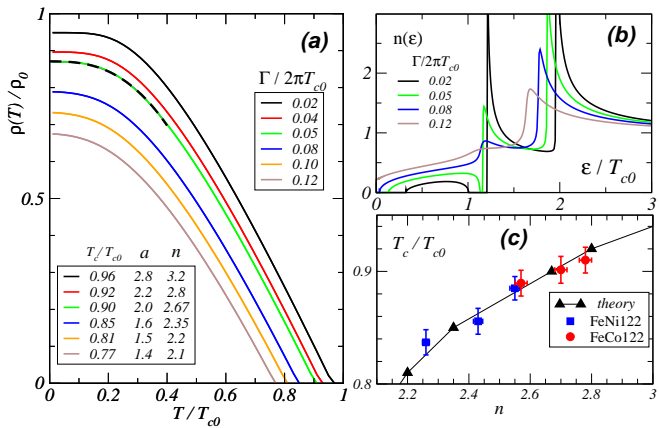}%
\caption{
(a) Superfluid density and (b) the density of states $n(\vare) = N(\vare)/N_f$,
computed for the $s^\pm$ state with sign-changing isotropic gaps
and strong interband impurity scattering, between the Born and unitary limits.
The dashed line in (a) is an example of a power-law fit
$\rho(T)/\rho_0 = \rho(0)/\rho_0 - a (T/T_{c0})^n$ for $0<T <0.4 \, T_{c0}$;
best fitting parameters for a given set of $\Gamma = n_{imp}/\pi N_f$ are listed in the table.
(b) As the impurity concentration $n_{imp} \sim \Gamma$)) increases,
the band of mid-gap states approaches the Fermi level and the exponent $n$ is reduced.
(c) $T_c$ vs. power $n$, from the theoretical model (triangles)
and experiment (squares and circles).
}
\label{fig4}
\end{figure}

The essential theoretical results are presented in Fig.~\ref{fig4} while the full
calculations will be published elsewhere \cite{vorontsov10}. The best agreement
with the experiment is obtained for two isotropic gaps,
$\Delta_2 \approx -0.6 \Delta_1$, strong inter-band scattering $\delta v=0.9$
and phase shift $\delta = 60^\circ$ between the Born ($\delta \to 0$) and
unitary limits ($\delta \to 90^\circ$). The calculated $\rho(T)$ was fitted to the
power-law, $\rho(T)/\rho_0 \approx \rho(0)/\rho_0 - a \left( {T}/{T_{c0}}\right)^n$,
which is directly related to the penetration depth,
$\Del\lambda(T)/\lambda_0 \approx a' (T/T_c)^n$, with
$\rho_0$ and $\lambda_0$ being the $T=0$ superfluid density and penetration depth
in the clean system and $a' = (a/2) [T_c/T_{c0}]^n [\rho_0/\rho(0)]^{3/2}$.
We find that with an increase in $\Gamma$, the power $n$ decreases from $n \gtrsim
3$ to $n \approx 2$ (see Fig.~\ref{fig4}(a)),
which is in perfect agreement with experiment. The values of $n$ depend
{\it sensitively} on the structure of the
low-energy density of states, which is shown in Fig.~\ref{fig4}(b).
The intermediate strength of scatterers
is important for creation of a {\it small} band of mid-gap states separated
from the continuum. As the disorder is increased, these states close the gap in
the spectrum and gradually increase in magnitude, driving the low-temperature
power-law dependence from exponential-like, $n > 3$, to $n \approx 2$.
This behavior of $n$ is largely independent of the details of the model,
whereas $a'$ can slightly increase or decrease depending on the
different ratios of the gaps on two Fermi surfaces and different impurity parameters.

Finally, in Fig.~\ref{fig4}(c) we show the central result of our study:
the correlation between $T_c$ and $n$.
Note that these two quantities are
obtained essentially independently of each other. Assuming that the unirradiated samples
have some disorder due to doping, and scaling $T^\un_c$
to lie on the theoretical curve, we find that the $T_c(B_{\phi})$ of the
irradiated samples also follows this curve. The assumption of similarity between
doping and radiation-induced disorder, implied in this comparison,
while not unreasonable, deserves further scrutiny.

In summary, we determined the effect of irradiation on $\lambda(T)$ and
demonstrated that the disorder-induced reduction of the power-law exponent
$2<n<3$ is naturally explained in terms of the isotropic extended $s$-wave
state \cite{Chubukov2009,Mazin2009} with pair-breaking interband
scattering\cite{Chubukov2009,Gordon2010,Glatz2010}. We have also considered
models for nodal states, but they showed the opposite trends: increase of $n$
with disorder in the interval $1<n<2$, and thus can be excluded. Taken together
with reports of fully gapped states from thermal conductivity
\cite{Makariy2010} and angle resolved photoemission spectroscopy
\cite{Terashima2009}, our results present the most convincing case, up to date,
in favor of the extended $s^\pm$ pairing symmetry having a nodeless order parameter
in the optimally doped 122 system.

The picture of strong pair-breaking scattering is also consistent with recent
proposals of the universal behavior in the thermal and electromagnetic
responses of iron-based superconductors \cite{Bud'ko2009,Gordon2010,Kogan2010}.
Nonetheless, we should note that nodal states may still exist in P-containing
compounds \cite{Fletcher2009,Hicks2009,Shishido2010} or along the $c$-axis of
heavily overdoped Ba122 pnictides \cite{Martin2010}.

We thank V. Kogan, J. Schmalian, A. Chubukov, I. Mazin, P. Hirschfeld and A. Koshelev for useful
discussions. This work was supported by the U.S. Department of Energy, Office of Basic Energy Sciences,
Division of Materials Sciences and Engineering under contract No. DE-AC02-07CH11358 (Ames National
Laboratory) and contract No. DE-AC02-06CH11357 (Argonne National Laboratory). The heavy ion irradiation
was performed at the ATLAS facility at Argonne. R.P. acknowledges support from the Alfred P. Sloan
Foundation.


\begin{thebibliography}{99}

\bibitem{Kamihara2008}Y.~Kamihara \emph{et~al.}, J. Am. Chem. Soc. \textbf{130}, 3296 (2008).

\bibitem{Ishida2009}K.~Ishida \emph{et~al.}, J. Phys. Soc. Jpn. \textbf{78}, 062001 (2009).

\bibitem{Chubukov2009}A.~V.~Chubukov, Physica C \textbf{469}, 640 (2009).

\bibitem{Mazin2009}I.~I.~Mazin and J.~Schmalian, Physica C \textbf{469}, 614 (2009).

\bibitem{Mazin2008}I.~I.~Mazin \emph{et~al.}, Phys. Rev. Lett. \textbf{101}, 057003 (2008).

\bibitem{Kuroki08}K.~Kuroki \emph{et~al.}, Phys. Rev. Lett. \textbf{101}, 087004 (2008).

\bibitem{Barzykin08}V.~Barzykin and L.~P.~Gorkov, JETP Letters \textbf{88},  131  (2008).

\bibitem{SeoBernevig08}K.~Seo \emph{et~al.}, Phys. Rev. Lett. \textbf{101},  206404 (2008).

\bibitem{Maier09gap}T.~A.~Maier \emph{et~al.}, Phys. Rev. B \textbf{79},  224510  (2009).

\bibitem{Graser09}S.~Graser \emph{et~al.}, New J. Phys. \textbf{11},  025016  (2009).

\bibitem{Chubukov09nodes}A.~V.~Chubukov \emph{et~al.}, Phys. Rev. B \textbf{80},  140515(R) (2009).

\bibitem{Thomale09nodes}R.~Thomale \emph{et~al.}, Phys. Rev. B \textbf{80},  180505(R)  (2009).

\bibitem{Tinkham}M.~Tinkham, {\it Introduction to superconductivity} (Dover Publications, Inc., 1996).

\bibitem{Bang2009} Y.~Bang, Euro. Phys. Lett. \textbf{86}, 47001 (2009).

\bibitem{Vorontsov2009}A.~B.~Vorontsov \emph{et~al.}, Phys. Rev. B \textbf{79}, 140507 (2009).

\bibitem{Gordon2010}R.~T.~Gordon \emph{et~al.}, arXiv:0912.5346v1 (2010).

\bibitem{Hirschfeld1993}P.~J.~Hirschfeld and N.~Goldenfeld, Phys. Rev. B \textbf{48}, 4219 (1993).

\bibitem{Gordon2009}R.~T.~Gordon \emph{et~al.}, Phys. Rev. Lett. \textbf{102}, 127004 (2009).

\bibitem{Gordon2009a}R.~T.~Gordon \emph{et~al.}, Phys. Rev. B \textbf{79}, 100506(R) (2009).

\bibitem{Hashimoto2009a}K.~Hashimoto  \emph{et~al.}, Phys. Rev. Lett. \textbf{102}, 207001 (2009).

\bibitem{Martin2009a}C.~Martin \emph{et~al.}, Phys. Rev. B \textbf{80}, 020501R (2009)

\bibitem{Martin2009}C.~Martin \emph{et~al.}, Phys. Rev. Lett. \textbf{102}, 247002 (2009).

\bibitem{Martin2010}C.~Martin \emph{et~al.}, Phys. Rev. B \textbf{81}, 060505(R) (2010).

\bibitem{Luan2010} L.~Luan \emph{et al.}, Phys. Rev. B \textbf{81}, 100501(R) (2010).

\bibitem{Kim2010}H. Kim \emph{et~al.}, arXiv:1001.2042 (2010).

\bibitem{Jin1993}S.~Jin, ed., {\it Processing and Properties of High-$T_c$
  Superconductors} (World Scientific, 1993).

\bibitem{Blatter1994}G.~Blatter \emph{et~al.}, Rev. Mod. Phys. \textbf{66}, 1125 (1994).

\bibitem{Zhu1993}Y.~Zhu \emph{et~al.}, Phys. Rev. B \textbf{48}, 6436 (1993).

\bibitem{Woods1998}S.~I.~Woods \emph{et~al.}, Phys. Rev. B \textbf{58}, 8800 (1998).

\bibitem{Canfield2009}P.~C.~Canfield \emph{et~al.}, Phys. Rev. B \textbf{80}, 060501(R) (2009).

\bibitem{Ni2008}N~.Ni \emph{et~al.}, Phys. Rev. B \textbf{78}, 214515 (2008).

\bibitem{Degrift1975}C.~T.~Van Degrift, Rev. Sci. Instrum. \textbf{46}, 599 (1975).

\bibitem{Prozorov2000}R.~Prozorov \emph{et~al.}, Phys. Rev. B \textbf{62}, 115 (2000).

\bibitem{Prozorov2000a}R.~Prozorov \emph{et~al.}, Appl. Phys. Lett. \textbf{77}, 4202 (2000).

\bibitem{Prozorov2006} R.~Prozorov and R.~W.~Giannetta, Supercond. Sci. Technol. \textbf{19}, R41 (2006).

\bibitem{Prozorov2010}R.~Prozorov \emph{et~al.}, Phys. Rev. B \textbf{81}, 094509 (2010).

\bibitem{mishra09}V.~Mishra, \emph{et~al.}, Phys. Rev. B \textbf{80}, 224525 (2009).

\bibitem{vorontsov10} A.~B.~Vorontsov, (unpublished).

\bibitem{Glatz2010} A.~Glatz and A.~E.~Koshelev, arXiv:1002.0363 (2010).

\bibitem{Makariy2010}M.~A.~Tanatar \emph{et~al.}, Phys. Rev. Lett. \textbf{104}, 067002 (2010).

\bibitem{Terashima2009}K.~Terashima \emph{et~al.}, PNAS \textbf{106}, 7330 (2009).

\bibitem{Bud'ko2009} S.~.L.~Bud'ko \emph{et~al.}, Phys. Rev. B \textbf{79}, 220516(R) (2009).

\bibitem{Kogan2010}V.~G.~Kogan, arXiv:1002.3390 (2010).

\bibitem{Fletcher2009}J.~D.~Fletcher \emph{et~al.}, Phys. Rev. Lett. \textbf{102}, 147001 (2009).

\bibitem{Hicks2009}C.~W.~Hicks \emph{et~al.}, Phys. Rev. Lett. \textbf{103}, 127003 (2009).

\bibitem{Shishido2010}H.~Shishido \emph{et~al.}, Phys. Rev. Lett. \textbf{104}, 057008 (2010).


\end{thebibliography}
\end{document}